\documentclass{mn2e}
\usepackage{psfig}
\usepackage{amsmath}
\usepackage[totalwidth=510pt, totalheight=700pt]{geometry}
\catcode`\@=11
\def\gsim{\ifmmode{\,\mathrel{\mathpalette\@versim>\,}}
    \else{$\,\mathrel{\mathpalette\@versim>}\,$}\fi}
\def\lsim{\ifmmode{\,\mathrel{\mathpalette\@versim<\,}}
    \else{$\,\mathrel{\mathpalette\@versim<}\,$}\fi}
\def\@versim#1#2{\lower 2.9truept \vbox{\baselineskip 0pt \lineskip
    0.5truept \ialign{$\m@th#1\hfil##\hfil$\crcr#2\crcr\sim\crcr}}}
\catcode`\@=12  
\renewcommand\[{\begin{equation}}
\renewcommand\]{\end{equation}}
\def\fracj#1#2{\textstyle{#1\over#2}}
\def\erf{\mathop{\rm erf}}
\def\e{{\rm e}}
\def\d{{\rm d}}
\def\i{\relax\ifmmode{\rm i}\else\char16\fi}

\def\ampl{P_0 \omega_0}
\def\av#1{\langle#1\rangle}

\def\dt{\Delta t}

\def\dtonefour{\Delta t_{1.4}}
\def\dtb{\Delta t_{\rm b}}
\def\deltat14{\Delta t_{1.4}}
\def\dtrad{\Delta t_{\rm rad}}
\def\fdt{F_{\dt}}
\def\fb{f_{\rm b}}
\def\few{\rm few}
\def\trfdt{\hat{F}_{\dt}}
\def\lx{L_{\rm X}}
\def\lb{L_{\rm B}}
\def\lstar{L^{\star}}
\def\lbstar{L_{\rm B}^{\star}}
\def\lmech{L_{\rm m}}
\def\nlm{\tilde{L}_{{\rm m}}} 
\def\elldt{\ell_{\rm \dt}}

\def\lonefour{L_{1.4}}
\def\lrad{L_{\rm rad}}
\def\p14{P_{1.4}}

\def\k14{\kappa_{1.4}}
\def\krad{\kappa_{\rm rad}}
\def\omegamin{\omega_{\rm min}}
\def\omegadt{\omega_{\dt}}
\def\omegamax{\omega_{\rm max}}
\def\prob{\mathcal{P}}

\def\mbh{M_{\rm BH}}
\def\mb{M_{\rm B}}
\def\phix{\Psi}
\def\rg{r_{\rm g}}
\def\sigdt{\sigma^2_{\dt}}
\def\sigx{\sigma^2_{\rm X}}
\def\sigb{\sigma^2_{\rm b}}
\def\sigrad{\sigma^2_{\rm rad}}
\def\sig14{\sigma^2_{\rm 1.4}}
\def\sigstar{{\sigma}^2_{\star}}

\def\tmin{t_{\rm min}}
\def\tmax{t_{\rm max}}

\def\tcs{t_{\rm c_{\rm s}}}

\def\trec{t_{\rm rec}}
\def\tsync{t_{\rm sync}}

\def\eb{E_{\rm b}}

\def\cmm3{\,{\rm cm}^{-3}}
\def\msun{M_{\odot}}

\def\gauss{\,{\rm G}}
\def\myr{\,{\rm Myr}}
\def\yr{\,{\rm yr}}
\def\dayunit{\,{\rm day}}

\def\hz{\,{\rm Hz}}
\def\mhz{\,{\rm MHz}}
\def\ghz{\,{\rm GHz}}
\def\sm1{\,{\rm s}^{-1}}

\def\km{\,{\rm km}}
\def\mpc{\,{\rm Mpc}}
\def\ergs{\,{\rm erg}}

\title[AGN variability and  cooling flows]{Time variability of AGN and heating of cooling flows}

\author[C. Nipoti and J. Binney]{Carlo Nipoti\thanks{E-mail: nipoti@thphys.ox.ac.uk} and James 
Binney
\\
Theoretical Physics, Oxford University,   1 Keble Road, Oxford OX1 3NP, UK\\}

\begin{document}

\date{Accepted 2005  May 3; in original form 2004 October 8}

\pagerange{\pageref{firstpage}--\pageref{lastpage}} \pubyear{2005}

\maketitle

\label{firstpage}

\begin{abstract}
There is increasing evidence that AGN feedback is important in the
energetics of cooling flows in galaxies and galaxy clusters. It is possible
that in most cooling-flow clusters radiative losses from the thermal plasma
are balanced, over the cluster lifetime, by mechanical heating from the
central radio source. We investigate the implications of the variability of
AGN mechanical luminosity $\lmech$ on observations of cooling flows and
radio galaxies in general.

It is natural to assume that $\ell=\ln(\lmech/\lx)$ is a Gaussian process.
Then $\lmech$ will be log-normally distributed at fixed cooling luminosity
$\lx$, and the variance in a measure of $\lmech$ will increase with the
time-resolution of the measure. We test the consistency of these predictions
with existing data for cooling flows and radio galaxies. These tests hinge
on the power spectrum $P(\omega)$ of the Gaussian process $\ell(t)$. General
considerations suggest that $P$ is a power law $P\sim\omega^{-\beta}$.
Long-term monitoring of Seyfert galaxies combined with estimates of the duty
cycle of quasars imply that $\beta\simeq1$, which corresponds to flicker
noise.  The power spectra of
microquasars have similar values of $\beta$.  We combine a sample of sources
in cooling flows that have cavities with the assumption that the average
mechanical luminosity of the AGN equals the cooling flow's X-ray luminosity.
Given that the mechanical luminosities are characterized by flicker noise, we
find that their spectral amplitudes $\omega P(\omega)$ lie between the
estimated amplitudes of quasars and the measured values for the radio
luminosities of microquasars. 

The model together with the observation that powerful radio galaxies
lie within a narrow range in optical luminosity, predicts the
luminosity function of radio galaxies. Both the shape and the
normalization of the predicted function are in agreement with
observations. Forthcoming radio surveys will test the prediction
that the luminosity function turns over at about the smallest
luminosities so far probed.
\end{abstract}

\section{Introduction}

In the cores of the majority of rich galaxy clusters the radiative
cooling time of the hot, X-ray emitting intracluster medium (ICM) is
significantly shorter than the Hubble time. In the absence of heat
sources, the gas will cool and flow towards the bottom of the cluster
potential well: hence the name `cooling-flow' clusters.  In fact,
recent X-ray observations -- in particular {\it XMM-Newton}
high-resolution spectroscopy -- have ruled out the presence of
significant cooling flows (Kaastra et al. 2001; Molendi \& Pizzolato
2001; Peterson et al. 2001; Tamura et al. 2001; Kaastra et al. 2004,
and references therein): the radiating gas in the cores of these
clusters is heated somehow.

The most promising heating mechanism is feedback from the Active
Galactic Nucleus (AGN) hosted by the central galaxy of the cluster.
The basic idea of this AGN/cooling-flow scenario is that the cooling
ICM in the cluster core is heated by outbursts from the central AGN:
an unsteady equilibrium may be achieved, in the sense that the energy
input from the AGN balances the radiative losses of the ICM over the
lifetime of the cluster.  Different models have been proposed in which
the AGN heats the thermal plasma with either mechanical (Binney \&
Tabor 1993; Tabor \& Binney 1995) or radiative feedback (Ciotti \&
Ostriker 1997, 2001; Ostriker \& Ciotti 2005). Although both processes
might be effective to some extent and possibly cooperate, here we
focus on mechanical heating, whose relevance to many observed systems
is apparent (e.g. Blanton et al. 2004 and Omma \& Binney 2004 and references
therein).

 The cooling luminosity $\lx$ can be determined directly from X-ray
observations (e.g. Peres et al.  1998).  By contrast, the mechanical
 power $L_{\rm m}$ of the radio sources is not only much harder to
 determine observationally (e.g. Bicknell 1995; Eilek 2004), but it is
 expected to be time-variable (e.g. Ulrich, Maraschi
\& Urry 1997). If $L_{\rm m}$ varies with time, how can it be compared with
the cooling luminosity in observed systems? This paper addresses this question.

\section{Conceptual framework}

The mechanical luminosity of an AGN is  expected to fluctuate on  time-scales
that range from months to several gigayears, but from the point of view of
the symbiosis between the AGN and its cooling flow, only  time-scales in
excess of $\sim10\myr$ are of interest. It is clearly impossible to probe
such  time-scales by monitoring individual AGN. Instead we must draw
inferences by studying the demography of AGN and cooling flows. Such studies
are still in their infancy, but it is nevertheless worthwhile to develop
a methodology that they can employ, and to see what  conclusions this
methodology and currently available data lead us to.  

\begin{figure}
\centerline{\psfig{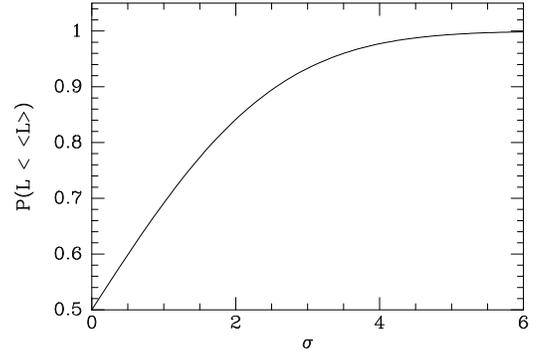}}
\caption{The probability for a log-normal distribution that a single
measurement yields a value smaller than the expectation value.\label{plessfig}}
\end{figure}

Cooling flows vary in size from those associated with large clusters of
galaxies, such as Abell 1795, to those associated with individual
elliptical galaxies, such as NGC 4472. We start by scaling out the
variations in the mechanical luminosities $L_{\rm m}$ of AGN that
simply reflect variations in size rather than temporal variability, by
focusing on the dimensionless variable
\[
\tilde L_{\rm m}=L_{\rm m}/\lx
\]
 where $\lx$ is the X-ray luminosity of the cooling flow, which is expected
to vary at most slowly and weakly with time. The logarithm of $\tilde L_{\rm
m}$,
 \[
\ell\equiv\ln\left( {L_{\rm m}/ \lx}\right),
\]
 is a
continuous random variable defined on the interval $(-\infty,\infty)$.
Throughout physics one models such a variable through the Fourier
representation 
 \[
\ell(t)=\int\d\omega\, A_\omega\e^{\i\omega t}.
\]
 The complex numbers $A_\omega$ are random variables. The simplest
conjecture is that these variables are statistically independent of one
another, and have phases that are uniformly distributed in $(0,2\pi)$. In
this case $\ell(t)$ is a Gaussian random process, and it is completely
characterized by the power spectrum $P(\omega)\equiv \av{|A_\omega|^2}$. The
probability density of the luminosity at any given time $t$ is given by
\begin{equation}\label{eqln}
\prob(\nlm) \d \nlm  = {1 \over {\sqrt{2\pi} \sigma}} {\exp\left[{-{( \ell
        -{\av{\ell}} )^2 \over 2\sigma^2}}\right]} \d \ell,
\end{equation} 
 where $\sigma^2 = \av{(\ell-{\av{\ell}})^2}$ is the variance of
$\ell$.  It is easy to show that with this model the expectation value of
the $n$th power of $\tilde L_{\rm m}$ is
 \begin{equation}\label{eqavln}
\av{\tilde L_{\rm m}^n} = \exp\left({n{\av{\ell}} +n^2{\sigma^2 \over 2}}\right).
\end{equation} 
 The appearance of $\sigma^2$ in the exponential signals the
skewness of the probability distribution (\ref{eqln}). The larger the value
of $n$, the more important this skewness is in the sense that rare excursions
to high $\tilde L_{\rm m}$ more strongly dominate $\av{\nlm^n}$. A
measure of the importance of skewness is the probability that an individual
measurement of $\nlm$ will return a value smaller than
$\av{\tilde L_{\rm m}}$. One can readily show that
 \[
P(\tilde L_{\rm m}<\av{\tilde L_{\rm m}})=
\fracj12[1+\erf(\sigma/2^{3/2})].
\]
 Figure \ref{plessfig} plots this function of $\sigma$.
When $\sigma\gsim3$, there is a high probability of
finding the luminosity to be smaller than its mean value because much of the
total energy output comes during infrequent outbursts.

We are interested in the properties of the light-curve observed with a
finite time-resolution, i.e., averaged over a given time $\dt$. 
We define this `coarse-grained' light-curve as
\begin{equation} 
\elldt(t)= \int_{0}^{\infty}\d t'\, \fdt(|t'-t|) \ell(t'),
\end{equation}
where the filter function $\fdt(t)$ satisfies $\int\d t\, \fdt=1$ and is
non--zero in a region of width $\dt$ around the origin. $\elldt(t)$ has
average
\begin{equation} 
\av{\elldt}=  \lim_{t \to \infty} {1 \over t}  \int_0^{t}\d t'\, \elldt(t')
 = \av{\ell},
\end{equation}
and variance\footnote{Note the strict similarity between the variance
  of the coarse-grained light-curve $\sigdt$ and the mass variance
  $\sigma^2_{M}$ usually adopted in cosmology to describe mass density
  fluctuations on mass scales larger than $M$.}
 \begin{eqnarray}\label{eqsigdt}
\sigdt=\lim_{t \to \infty} {1 \over t} \int_0^{t}\d t\, (\elldt(t')-\av{\ell})^2 \nonumber\\
= \int_{0}^{\infty}\d\omega\, \trfdt^2(\omega) P(\omega)
\end{eqnarray}
 where $\trfdt(\omega)$ is the Fourier transform of the filter
$\fdt(t)$. Approximating the window function ${\trfdt}^2(\omega)$ with
the Heaviside step function $\theta(1-\omega/\omegadt)$, where
${\omegadt}=2\pi/\dt$, we get
\begin{equation}\label{eqsigapp} 
\sigdt \simeq  \int_{0}^{\omega_{\dt}}\d\omega\, P(\omega).
\end{equation}
In words, contributions to the variance on time-scales shorter than
$\dt$ vanish in the averaging process. Thus, for any power spectrum
$P(\omega)$, $\sigdt$ is a non--increasing function of $\dt$, with
$\lim_{\dt \to \infty} \sigdt = 0$ and $\lim_{\dt \to 0} \sigdt =
\sigma^2$: the variance is smaller when averaging over longer $\dt$.

\subsection{The power spectrum}

 An AGN involves an enormous dynamical range of time-scales, from the few
days characteristic of the dynamical time at the event horizon to the
hundreds of megayears required for the development of a cooling catastrophe
in the core of the cooling flow. In view of this great span of time-scales,
one might expect the power spectrum of the AGN to be a power law in
frequency.  So we assume that
$\ell(t)$ has a power spectrum in the form
 \begin{equation}\label{eqps}
P(\omega)=
\begin{cases} 
P_0 \left({\omega/ \omega_0}\right)^{-\beta} &  
\text{if $\omegamin \leq \omega \leq \omegamax$,}\\
0  & \text{otherwise,}
\end{cases}
\end{equation}
where $P_0$ is the power at some reference frequency $\omega_0$,
${\omegamin}=2\pi/{\tmax}$ and $\omegamax \equiv 2\pi/\tmin$. $\tmin$
and $\tmax$ are, respectively, the maximum and minimum time-scale of
variability.

There is no difficulty in principle in deriving equations for a general
power-law index $\beta$, but we shall argue that $\beta$ must lie close to
unity, and the discussion is greatly simplified if we specialize to
$\beta=1$ at this point. The case $\beta=1$ is associated with `flicker
noise', which empirically arises in diverse physical phenomena that range
from electrical noise, to wind speeds, to earthquake magnitudes (e.g.\
Press~1978; Milotti 2002).  In case of flicker noise the relation between
the variance and the power spectrum is
 \begin{equation}\label{eqoneoverf} 
{\sigma_{\dt}^2}=\ampl \ln{\omega_{\dt} \over \omegamin} = \ampl \ln{\tmax \over \dt}.
\end{equation}
The amplitude of flicker noise is conveniently characterized by the
dimensionless quantity $\omega P(\omega)$, which is independent of
$\omega$.

\subsubsection{Calibration with quasars}

The variability of X-ray emission from AGN has been studied on
time-scales shorter than $\sim10\yr$.  When the X-ray power
spectra\footnote{The authors consider the power spectrum of $L(t)$
normalized to $\av{L}^2$, while here we refer to the power spectrum of
$\ell(t)\equiv\ln L(t)$: however, the two have the same slope and
amplitude as long as the variance of $L(t)$ is small with respect to
$\av{L}^2$.} of Seyfert galaxies at frequencies $10^{-8}-10^{-5} \hz$
are fitted with a power-law, one finds $\beta\sim 1$ (for instance,
$\beta\simeq1.05^{+0.15}_{-0.20}$ for the Seyfert galaxy NGC\,4051),
with amplitudes $P(\omega)\sim 0.01-0.04$ (Uttley, McHardy \&
Papadakis 2002; McHardy et al. 2004).  Hence, the only power-law
indices that we can consistently adopt are $\beta\simeq1$.

There is a vast difference between the longest time-scales on which the
power spectra have been empirically determined, and the time-scales
$\ga10^7\yr$ of interest here. Hence to extrapolate blindly from the
observed $\la10\yr$ fluctuations out to $10^7\yr$ ones is a very
hazardous proceeding.  Fortunately, independent estimates of the duty
cycle of quasars provides a degree of reassurance. Ciotti \& Ostriker
(2001) define the duty cycle of a source with light-curve $L(t)$ as
\begin{eqnarray}\label{eqduty}
\delta &\equiv& { \av{L}^2 \over \av{L^2}}
=\e^{-\sigma^2},
\end{eqnarray}
 where the second equality follows from equation (\ref{eqavln}).  Various
arguments indicate that quasars, over the lifetime of their host galaxies
have duty cycles $\delta \sim 10^{-3}-10^{-2}$, with episodic lifetime
(i.e., duration of individual major outbursts) $\gsim 10^4 \yr$ (Haehnelt,
Natarajan \& Rees 1998; Yu \& Tremaine 2002; Martini \& Schneider 2003;
Haiman, Ciotti \& Ostriker 2004).  By equation (\ref{eqduty}), these duty
cycles require variances in the range $\sim 4.6-6.9$. Inserting these
variances and the time-scale range $10^4-10^{10} \yr$ into equation
(\ref{eqoneoverf}), we find that the associated flicker spectra have
amplitudes $\ampl \sim 0.3-0.5$. These values are larger by a factor
$\sim30$ than the amplitudes measured by Uttley et al.\ (2002) for Seyfert
galaxies at frequencies higher by $\ga10^6$. If $\beta$ differed
significantly from unity, we would expect the amplitudes to be many orders
of magnitude different, so we must have $\beta\simeq1$.

\subsubsection{Calibration from microquasars} 

Microquasars are radio-jet emitting X-ray binaries made up by a
stellar-mass black hole (BH) and a companion star, and seem in many
respects small-scale versions of AGN (Mirabel \& Rodriguez 1999).  It
is widely accepted that on small to intermediate scales these objects
are identical to AGN scaled down in both length- and time-scales by
the ratio of the black holes' Schwarzschild radii $\rg \equiv
2G\mbh/c^2$. Hence, we can study the temporal variability of
microquasars on human time-scales to get information about the
variability of AGN on time-scales much longer than those that can be
directly probed. It seems likely that the scaling holds at out to
time-scales\footnote{For the accretion disk of a $\sim 10^8 \msun$ BH,
the characteristic viscous time-scale at $\sim 10^3 \rg$ is roughly
$10^5 \yr$.  Longer time-scales correspond to larger radii, at which
AGN accretion disks might be truncated by self-gravity (e.g. Burderi,
King \& Szuszkiewicz 1998; Goodman 2003).}  $\sim10^5\yr$, which for
an AGN black-hole mass $10^7-10^8\msun$ corresponds to a readily
observed frequency $10^{-5}\hz$ in microquasars. Thus from studies of
microquasars we can estimate the amplitude of the AGN power spectrum
at frequencies corresponding to $10^5\yr$, and then use our knowledge
that $\beta$ cannot significantly differ from unity to extrapolate it
to frequencies lower by a factor $\sim100$.

Most studies of the low-frequency power spectrum of microquasars focus
on X-ray emission.  Reig, Papadakis \& Kylafis (2002) find that the
power spectrum of the microquasar GRS~1915+105 is well fitted
by a power law (equation~\ref{eqps}) with index $\beta\simeq1$ in the
frequency range $10^{-7}$ to $10^{-5} \hz$.  A power spectrum with
this slope, over the same frequency range, has been also found for the
X-ray light curve of Cygnus X-3 by Choudhury et al. (2004).

However, we are interested in the variability of the kinetic power of
the jet, which is expected to be traced by radio emission rather than
by X-rays. The radio variability of microquasars has not
been investigated as much as their X-ray variability.  A spectral
analysis was carried out by Fiedler at al. (1987) for the radio (2.25
and 8.3 GHz) light curve of the peculiar microquasar SS~433: they
found that its power spectrum is consistent with a power law with
index $\beta\simeq1$ at frequencies $10^{-7}$ to $10^{-6} \hz$ (see
also Revnivtsev et al. 2005).  Nipoti, Blundell \& Binney (2005) show
that at these low frequencies the {\it slopes} of the radio power
spectra of the microquasars Cygnus~X-3 and GRS~1915+105 are similar to
those found in X-rays.

The X-ray light-curve of GRS~1915+105 has power spectrum with
amplitude $\ampl \sim 0.02$ at frequencies $10^{-7}$ to $10^{-5}\hz$
(Reig et al.  2002). An amplitude smaller by a factor of $\sim 10$ has
been measured for the X-ray power spectrum of Cygnus~X-3 in the same
frequency range (Choudhury et al.~2004).  However, there is evidence
that the {\it amplitude} of the variability of the radio emission of
these powerful microquasars is larger than that of their X-ray
emission (Nipoti et al.~2005). The radio emission of Cygnus~X-3 (at
2.25 and 8.3 GHz) has been monitored by the GBI-NASA monitoring
program (Waltman et al.  1995, and references therein).  We find that
the daily-averaged 2.25-GHz radio light-curve of Cygnus~X-3 has
logarithmic variance $\sigma^2\simeq0.96$ over the period
$1982-2001$. If we adopt $\tmax=19 \yr$, $\dt={1~\dayunit}$ for the
boundaries of the frequency range sampled, equation~(\ref{eqoneoverf})
implies that the power spectrum of the radio emission has amplitude
$\ampl~\simeq~0.11$, about two orders of magnitude larger than that
found in X-rays.  Similarly, we estimated the amplitude of the radio
power spectrum of the microquasar GRS~1915+105, using its 2.25-GHz
light-curve obtained by the GBI monitoring program over about $6 \yr$
(see Foster et al. 1996).  We found that GRS~1915+105 has amplitude
$\ampl~\simeq~0.12$ at $2.25\,$GHz, also much larger than in X-rays.
Thus, observations of microquasars suggest $\ampl~\sim~0.1$ as a
plausible value of the amplitude of the very low-frequency power
spectrum of radio sources.  Given the substantial uncertainies
involved, this estimate is in remarkably good agreement with the
estimate $\ampl~\sim~0.4$ that we obtained by considering the duty
cycle of quasars.

\begin{figure}
\centerline{\psfig{file=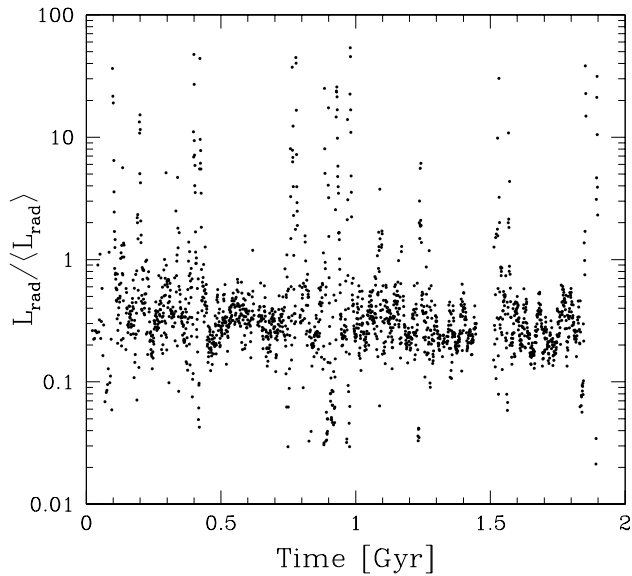,width=\hsize}}
\caption{Possible realization of the radio light-curve (normalized to
the averaged radio luminosity) of an AGN. The light-curve is obtained
from the daily-averaged 2.25 GHz radio light-curve of Cygnus~X-3 over
two decades by rescaling in time by a factor $10^8$.\label{pseudoFig}}
\end{figure}

These considerations suggest that we can obtain an impression of the
long-term variability of AGN by rescaling the radio-frequency light
curve of a microquasar.  Fig.~\ref{pseudoFig} is a plot of the
daily-averaged 2.25 GHz light-curve of Cygnus X-3 rescaled in time by
$10^8$. The light curve is characterized by short powerful outbursts,
alternating with periods of quiescence.

\section{Application to cooling flows}

\subsection{Systems with cavities}

In general, it is not feasible to measure directly the mechanical power of
radio sources. An exception is M87 -- the central galaxy of the Virgo
cluster -- for which high-resolution observations of the jet and the very
inner hot atmosphere are available (Owen, Eilek \& Kassim 2000; Di Matteo et
al. 2003).  In the case of the central radio sources in galaxy clusters
$L_{\rm m}$ can be estimated from depressions in the X-ray emission of the
surrounding ICM. These `cavities' are interpreted as bubbles produced in the
interaction between the jets of the radio source and the thermal plasma. The
energy transferred to the ICM by a bubble expanding adiabatically is (e.g.
Churazov et al.\ 2000, 2002)
 \begin{equation}\label{eqeb} 
\eb= {\gamma \over \gamma-1} p V,
\end{equation} 
 where $V$ is the volume of the bubble, $p$ is the pressure of the
surrounding ICM, and $\gamma$ is the adiabatic index of the gas in the
bubble. Thus, $\eb={5 \over 2}pV$ for a non-relativistic gas
($\gamma=\frac{5}{3}$), and $\eb=4pV$ for a relativistic fluid
($\gamma=\frac{4}{3}$).  Given a characteristic time $\dtb$ associated
with the bubble, an estimate of the average kinetic luminosity of the
AGN over $\dtb$ is then
\begin{equation}\label{eqlb}
\lmech={\eb \over \dtb}.
\end{equation}

The definition of $\dtb$ is controversial: if we consider $\dtb$ as
the {\it age} of the bubble, this is of order the time $\tcs$ required
for the cavity to rise at the speed of sound from the centre to its
current location (Gull \& Northover 1973; Churazov et
al. 2001). Alternatively, the age of the bubble can be defined as the
time required for the ICM to refill the displaced volume (McNamara et
al. 2000).  However, $\dtb$ is not necessarily the age of the bubble:
it could be the {\it recurrence time} $\trec$ of the radio source,
defined as the time elapsing between two subsequent outbursts in which
the jet blows bubbles. As we are looking for the characteristic time
to be associated with this measure of $\lmech$ (i.e., the time over which
we average in measuring from the bubbles), it is natural to identify
$\dtb$ with $\trec$, so in the rest of the paper we assume $\dtb
\equiv \trec$.

B\^{\i}rzan et al. (2004, hereafter B04) present radio and X-ray data
for a sample of 18 systems with cooling flows that show evidence of
cavities in their diffuse X-ray emission. The sample consists of 16
galaxy clusters, the galaxy group HCG~62, and the elliptical galaxy
M84.  The authors provide for each of these objects the cooling
luminosity ($\lx$), the monochromatic radio luminosity at 1.4 GHz
($\lonefour$), the integrated radio luminosity between 10 MHz and 5 GHz
($\lrad$), and measures of $pV$ for each cavity.  This sample is
suitable for comparison with the AGN/cooling flow model, because for
each source different measurements of the mechanical power (averaged
on different time-scales) are available.

\begin{figure}
\centerline{\psfig{file=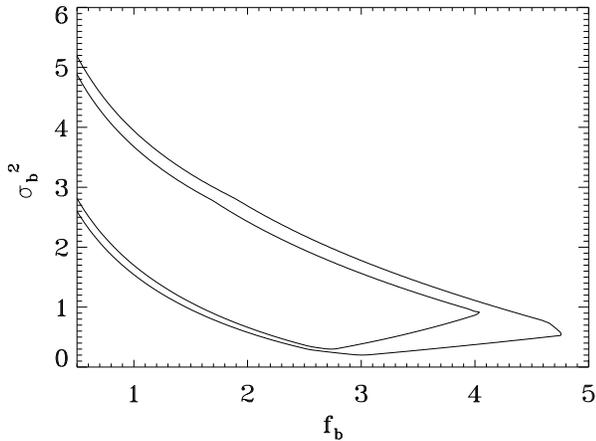,width=1.1\hsize}}
\caption{90 and 95 per cent confidence levels in the 
  space of the parameters $\fb=[\gamma/(\gamma-1)]/(\dtb/10^8 \yr)$
  and variance $\sigb$ for the distribution of $\nlm$ (measured from
  bubbles) for B04's sample of 18 objects with cavities. The null
  hypothesis is that the distribution is log-normal with $\av{\nlm}=
  1$.\label{fbfig}}
\end{figure}

From measures of $pV$ we derive $\lmech$ and $\nlm=\lmech/\lx$ using
equation~(\ref{eqlb}), given the time-scale $\dtb=\trec$. In general,
we do not know the recurrence time for individual sources.  In the
Perseus cluster, in which two generations of bubbles are observed
(Fabian et al. 2000), the recurrence time-scale is of the order of
$100 \myr$.  Estimating $\trec$ for other sources, in which just one generation
of cavities is observed, is problematic, but statistical arguments
suggest $\trec \sim \few \times 100 \myr$ as a typical value (B04).
So we assume that $\dtb$ is the same for all sources, and we
parameterize $\lmech$ as
\begin{equation}\label{eqfb}
\lmech=\fb {pV \over 10^8 \yr}, 
\end{equation}
where 
\begin{equation}
\fb \equiv {\gamma \over \gamma-1} \left({\dtb \over 10^8 
    \yr}\right)^{-1} .
\end{equation}
 We then use the Kolmogorov--Smirnov (KS) test to estimate how well the
measured values of $\nlm$ fit the predicted log-normal distribution
(\ref{eqln}) with $\av{\nlm}=1$. Fig.~\ref{fbfig} plots 
the 90 and 95 per cent confidence levels, derived from
the KS test, in the space of the parameters $\fb$ and $\sigma^2$.  For
instance, if we fix $\gamma/(\gamma-1)=4$, and allow $\dtb$ in the
range $1 \lsim (\dtb/10^8 \yr) \lsim 4$, the factor $\fb$ lies in the
interval $1 \lsim \fb \lsim 4$.  Such a range in $\fb$ corresponds to
values of the variance $0.2 \lsim \sigma^2 \lsim 3.9$ that give
distributions consistent with the data at the 95 per cent confidence
level.

\begin{figure}
\centerline{\psfig{file=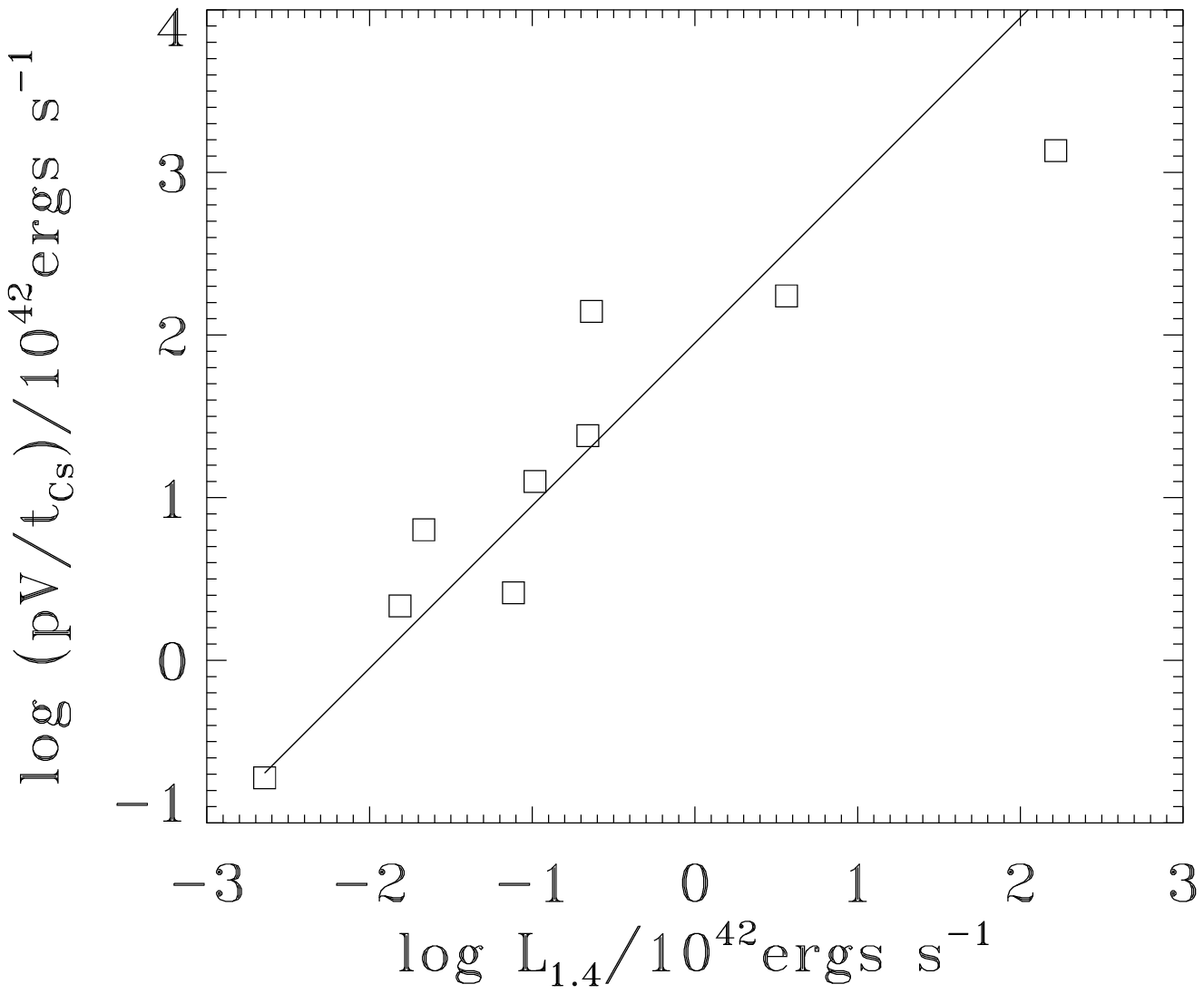,width=\hsize}}
\centerline{\psfig{file=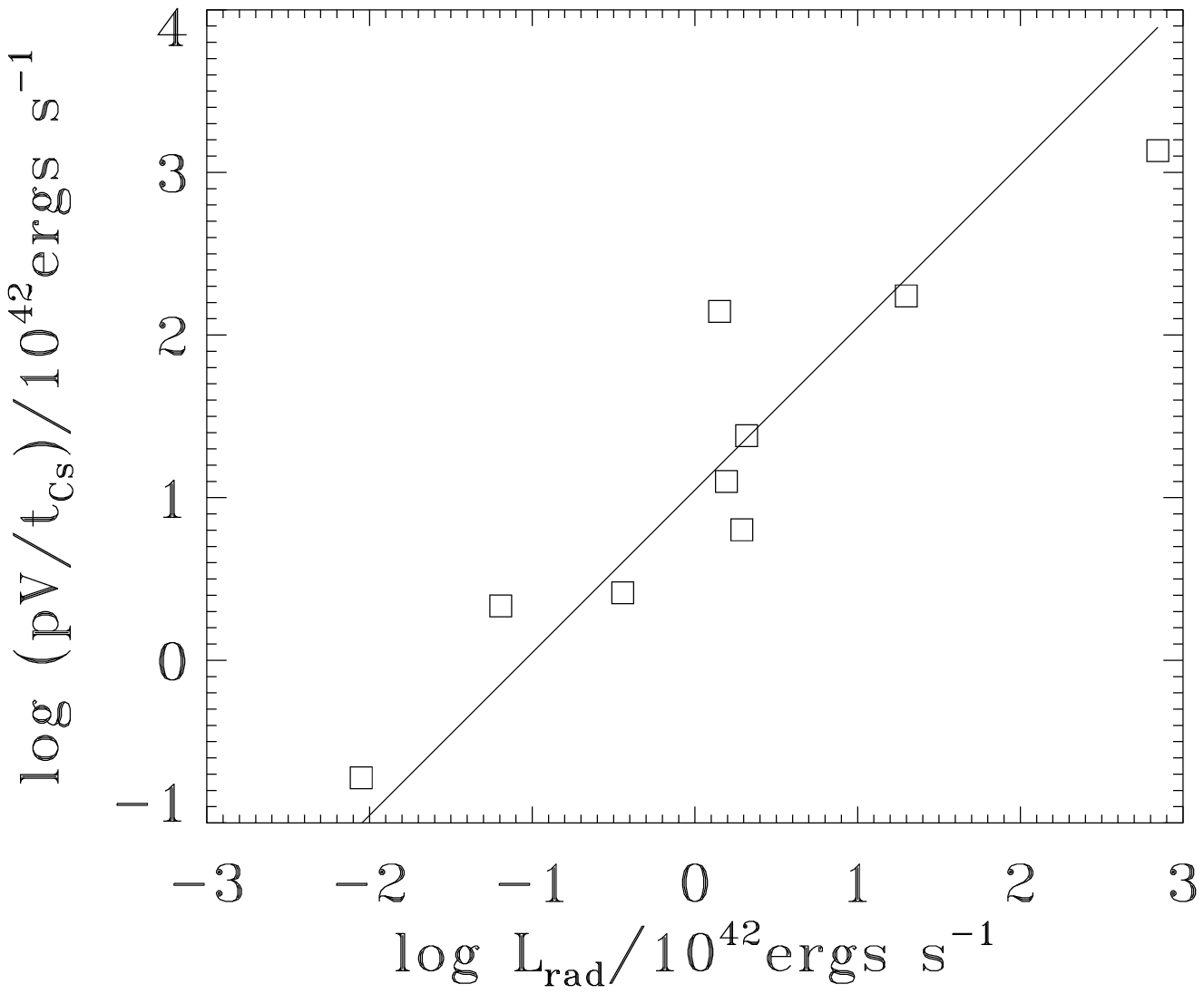,width=\hsize}}
\caption{The ratio $pV/\tcs$ as a function of the monochromatic 1.4 GHz
  luminosity (top) and the bolometric radio luminosity (bottom) for
  radio-filled cavities (data from B04).  The solid lines are
  the best--fitting linear relations: equations~(\ref{eqk14}) and
  (\ref{eqkrad}) with $\k14=89\gamma/(\gamma-1)$, and
  $\krad=11.1\gamma/(\gamma-1)$, respectively, assuming
  $\lmech=\eb/\tcs$.\label{konefourfig}}
\end{figure}

\subsection{Calibrating the radio luminosity}

The number of systems in which cavities have been accurately observed is
small, while the radio luminosities of thousands of radio galaxies have been
measured. It is clear that the radio power represents a very
small fraction of the kinetic power emitted by the AGN (e.g. Eilek 2004),
but the radio luminosity may be a useful indicator of the mechanical power.
Hence, much larger samples could be assembled if it were possible to
establish a relation between the mechanical and radio luminosities of sources.

Cavities in the X-ray emitting ICM are classified as radio-filled if
non-thermal radio emission at a given frequency coincides with the
X-ray-depressed area; otherwise they are called `ghost'
cavities\footnote{This classification depends on the frequency of the
radio emission. Cavities classified as ghosts on the basis of $\sim
\ghz$ radio data might appear as radio-filled at lower frequencies,
depending on the break frequency of the radio spectrum.  This has been
observed in Perseus by Fabian et al. (2002).}.  If reacceleration of
the synchrotron electrons is negligible, the age of the radio-filled
bubbles will be of the order of the synchrotron lifetime, or smaller,
while the age of the ghosts will be longer.  Thus, from radio-filled
cavities we can independently measure $L_{\rm m}\simeq\eb/\tcs$ and
$\lonefour$ and thus determine the relation
that holds between them.

The top panel of Fig.~\ref{konefourfig} plots data for the nine objects\footnote{We do not
include Abell 478, whose cavities are considered radio-filled by B04,
but are more likely to be already fading to ghosts, as suggested by
Sun et al. (2003). Also B04 point out the anomalous high ratio of
mechanical to radio power of this source.}  with {\it radio-filled}
cavities presented by B04. The values of $pV$ are from {\it Chandra}, and
$\tcs$ is from B04.  $\lonefour= \p14 \times 1.4 \ghz$, where
$\p14$ is the monochromatic $1.4\,$GHz power measured by  the VLA.
The straight line is the relation
\begin{equation}\label{eqk14} 
\lmech={\gamma\over\gamma-1}{pV\over\tcs}= \k14 \lonefour,
\end{equation}
with the best--fitting dimensionless conversion factor $\k14 \simeq 89 \gamma/(\gamma-1)$.

The lower panel of Fig.~\ref{konefourfig} shows the result of repeating this
exercise using the bolometric radio luminosity $L_{\rm rad}$ instead of the
monochromatic luminosity.  $\lrad$ is estimated by integrated between $10\,$MHz
and $5\,$GHz, assuming constant spectral index for each source (B04). 
The straight line is the relation 
\begin{equation}\label{eqkrad}
\lmech= \krad \lrad,
\end{equation} 
 with best--fitting $\krad \simeq 11.1 \gamma/(\gamma-1)$.  B04 found
 that a non-linear relation -- the mechanical luminosity scaling as
 $\lrad^{0.6
\pm 0.1}$ -- provides a better fit to their data (including Abell~478) than
the simple proportionality $\lmech \propto \lrad$ considered here.  However,
the sample is very small, and one can show that, excluding Abell~478 and the
peculiar radio source Cygnus~A -- the only FR II (Fanaroff \& Riley 1974)
source in the sample -- the slope of the best--fitting power-law is close to
1.  We think that a larger sample is needed to verify the existence of a
non--linear relation between kinetic and radio power.

\begin{figure}
\centerline{\psfig{file=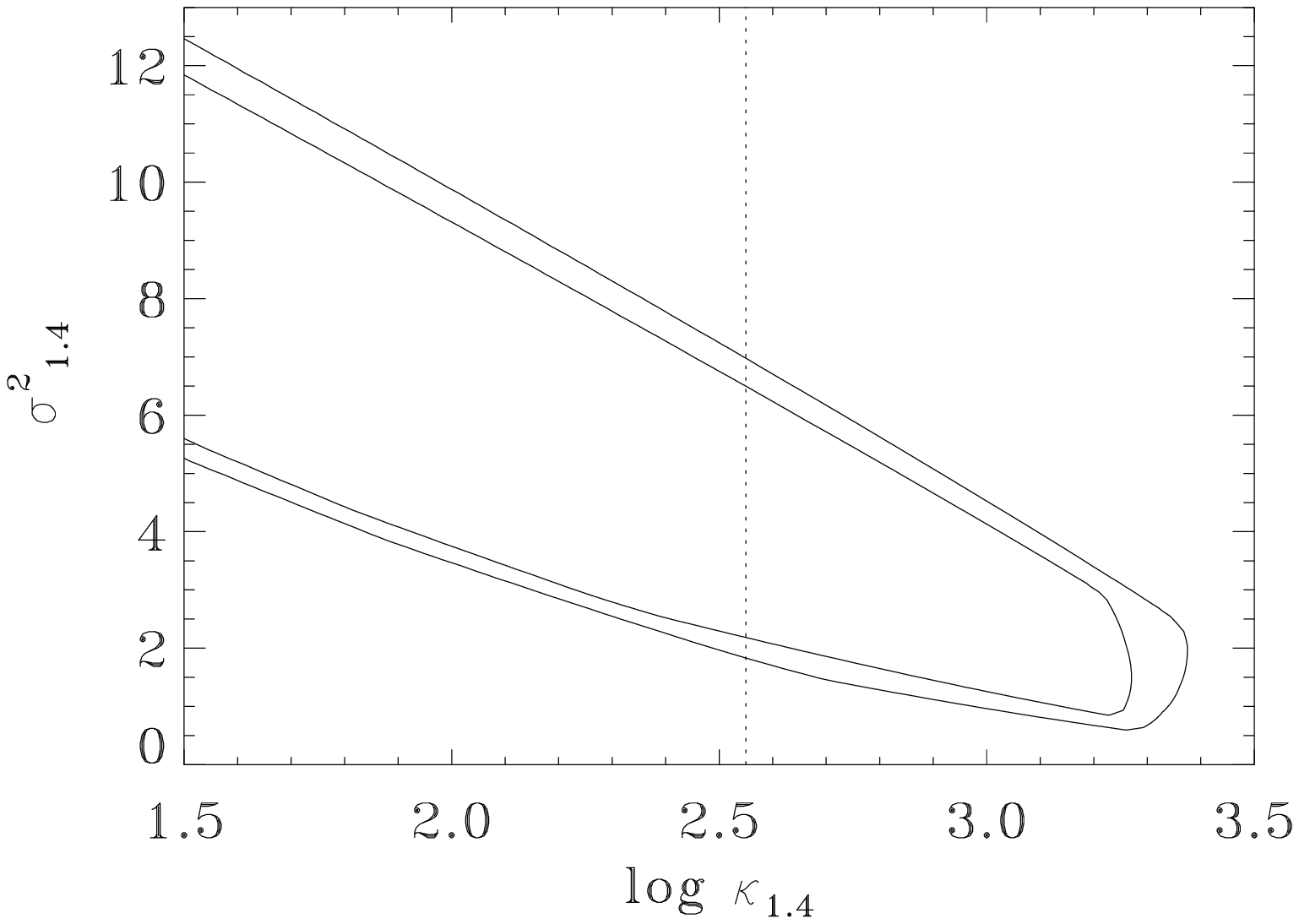,width=1.1\hsize}}
\centerline{\psfig{file=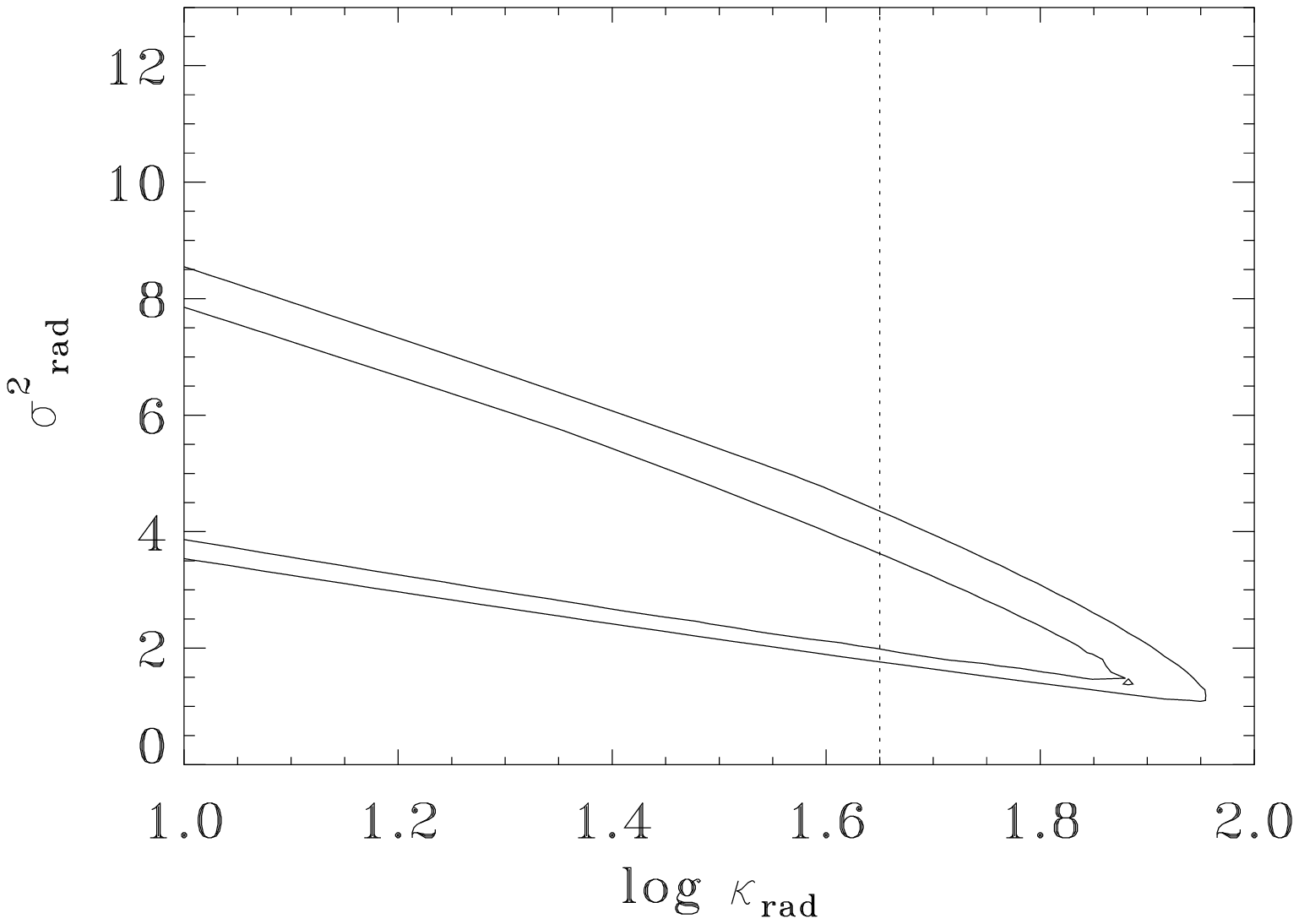,width=1.1\hsize}}
\caption{Top: 90 and 95 per cent confidence levels, in the space of
the parameters $\k14$ (equation~\ref{eqk14}) and variance $\sig14$ for
the distribution of $\nlm$ (measured from $\lonefour$) for B04's sample.
Bottom: the same as in top panel, but in the space $\krad$-$\sigrad$,
for the distribution of $\nlm$ measured from $\lrad$. The vertical
dotted lines correspond to the best--fitting values $\k14
\simeq 356$ and $\krad \simeq 44$ found for the radio-filled cavities,
assuming $\gamma/(\gamma-1)=4$. The null hypothesis is that the
distributions are log-normal with
$\av{\nlm}=1$.\label{k14sigfig}}
\end{figure}

We next investigate the constraints that can be placed on $\k14$ and $\krad$
by assuming that $\av{L_{\rm m}}=\lx$.  We use the whole sample presented by
B04, {\it without any distinction between radio-filled and ghost cavities},
and assume that the conversion coefficients $\k14$ and $\krad$ are the same
for all sources. We use the KS test to evaluate the consistency between the
values of $L_{\rm m}$ obtained using trial values of these coefficients and
the log-normal probability distribution (\ref{eqln}) with $\av{\nlm}=1$.
Fig.~\ref{k14sigfig} shows contours of 90 and 95 percent confidence levels
in the planes of $\sigma^2$ versus $\log(\k14)$ (top) or $\log(\krad)$
(bottom).  The dotted line in each panel shows the value of $\kappa$
determined above from Fig.~\ref{konefourfig} under the assumption that
$\gamma/(\gamma-1)=4$.  The data are consistent for a wide range of
parameters. Larger variances $\sigma^2$ are required for smaller
coefficients $\kappa$ because the current values of $L_{\rm m}$ are
proportional to $\kappa$, and if these are small the assumption $\av{\nlm}=1$
requires that significant energy is released during excursions to high
luminosity. For the values of $\kappa$ that we derived from radio-filled
bubbles, $\sigma_{\rm rad}^2\simeq3$ and $\sigma^2_{1.4}\simeq4$ are
favoured variances. These values are intermediate between the values
$\approx 1$ that we inferred from microquasars and $\simeq5$ that we
estimated from quasars.

\subsection{Dependence of $\sigma^2$ on time-scale}

As discussed in Section 2, the variance $\sigma^2$ associated with any
measurement of $L_{\rm m}$ should decrease with increasing timespan
over which that measurement effectively averages $L_{\rm m}$. Using
radio data we estimate the kinetic power averaged over the lifetime of
synchrotron-emitting relativistic electrons, which depends on the
radio frequency. A reference time-scale for synchrotron radiation at
frequency $\nu$ is the synchrotron cooling time
\begin{equation}\label{eqtsync}
\tsync \simeq 50   \left({B \over 10^{-5} \gauss}\right)^{-3/2} \left({\nu \over \ghz}\right)^{-1/2} \myr,
\end{equation} 
 where $B$ is the modulus of the magnetic field (e.g. Carilli et al.
 1991).  $\tsync$ depends strongly on $B$, which is usually poorly
 constrained in observed radio sources, and for fixed strength of the
 magnetic field decreases by about one order of magnitude from $\nu =
 70 \mhz$ to $\nu=5 \ghz$. Actually, $\tsync$ represents an upper
 limit on the lifetime of the radio-emitting electrons because
 adiabatic expansion losses are expected to be dominant over radiative
 losses (e.g. Blundell \& Rawlings 2000). As the effects of adiabatic
 expansion losses are difficult to quantify in general, in our
 treatment we simply assume that the averaging time-scale is $\lsim
 \tsync(\nu)$, when $\lmech$ is measured from radio luminosity at
 frequency $\nu$.

\begin{figure}
\centerline{\psfig{file=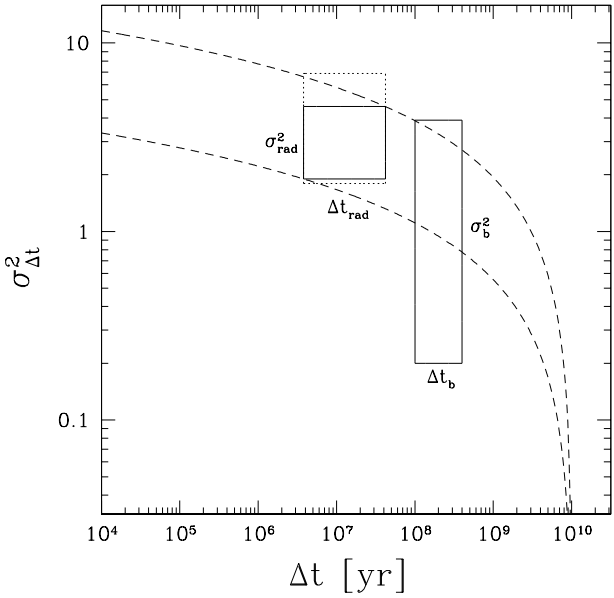,width=\hsize}}
\caption{Variance of the logarithm of the mechanical power of a radio
source, as a function of $\dt$, the time over which the mechanical
power is averaged. The two solid boxes define the loci of the diagram
with pairs ($\dtb$,$\sigb$) and ($\dtrad$,$\sigrad$) derived for B04's
sample from observations of X-ray cavities and bolometric radio
luminosity, respectively. The dashed curves are the upper and lower
limit of the family of functions $\sigdt$ predicted for a $1/f$ power
spectrum, which are consistent with the data. The dotted box refers to
pairs ($\dtonefour$,$\sig14$) derived from monochromatic 1.4 GHz data
for the same sample.\label{obsdtfig}}
\end{figure}

Fig.~\ref{obsdtfig} plots the three ranges of acceptable values of
$\sigma^2$ that we have derived above against the characteristic
averaging time $\Delta t$ associated with each measurement.  For the
mechanical luminosity measured from the bubbles, we have $0.2 \lsim
\sigb \lsim 3.9$, and we adopt $100 \lsim \dtb \lsim 400 \myr$ as the 
range for the recurrence time (see Section~3.1).  
For the mechanical luminosity derived from 1.4 GHz radio data, $1.8
\lsim \sig14 \lsim 6.9$ and $3 \lsim \dtonefour \lsim 36 \myr$, where the
range in $\dtonefour=\tsync(1.4 \ghz)$ has been computed from
equation~(\ref{eqtsync}) for $B$ in the range $(1-5) \times 10^{-5}
\gauss$ -- consistent with minimum-energy magnetic fields measured in
the radio lobes of 3C~218 (Hydra~A; Taylor et. al 1990), 3C~84
(Perseus; Fabian et al. 2002), and 3C~405 (Cygnus~A; Carilli et al.
1991). Considering the bolometric radio luminosity, we get more
stringent limits on the variance, $1.9 \lsim \sigrad \lsim 4.6$, while
the corresponding $\dtrad$ spans the same range as $\dtonefour$,
because in this case the bolometric luminosity is measured from $1.4
\ghz$ radiation. We assume that the relevant limits on the variance
are those found for $\sigrad$, because the spread in the slopes of the
radio spectra is likely to contribute to $\sig14$.

The dashed curves in Fig.~\ref{obsdtfig} show the dependence of
$\sigma^2$ upon $\Delta t$ that is expected for flicker noise spectra
(equation~\ref{eqsigapp}) with two different amplitudes, $\ampl=0.24$
and $0.84$. The model and the data are consistent for amplitudes that
lie within this range, which overlaps the interval between the
amplitudes of the radio spectra of microquasars and those inferred
from the duty cycle of quasars.  Note that for $\dtrad$ we used the
synchrotron cooling time, which is an upper limit. A shorter $\dtrad$
would result in lower value of $\ampl$: for instance, reducing
$\dtrad$ by a factor of 10 we get $0.19 \lsim \ampl \lsim 0.59$.

\section{Application to the radio-galaxy  luminosity function}

\begin{figure}
  \centerline{\psfig{file=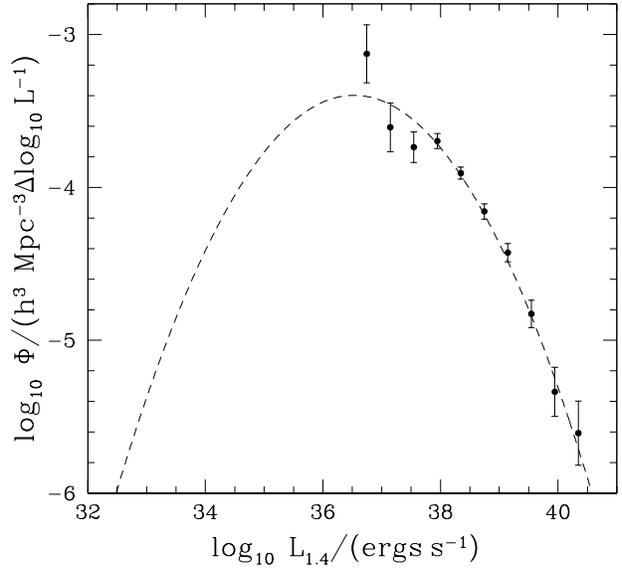,width=\hsize}}
\caption{Local radio luminosity function of early-type
galaxies. Points with error bars are data from the FIRST-2dFGRS
(Magliocchetti et al. 2002). The dashed line is the best--fitting
log-normal distribution (equation~\ref{eqrlfln}).\label{agnrlffig}  }
\end{figure}

In the previous section we have shown that, under plausible
assumptions, observed data of central radio galaxies in cooling-flow
clusters are consistent with the hypothesis that AGN heating balances
radiative cooling. Here we discuss the
consequences of extending the considered picture to cooling-flow
systems in general.  In particular, applying our model to luminous
early-type galaxies leads to predictions on the local radio luminosity
function.

Radio galaxies, though spanning several orders of magnitude in radio
power, represent a very homogeneous family of objects when the
properties of their host stellar systems are considered, being
early-type galaxies with optical luminosity larger than $\lstar$. For
instance, Magliocchetti et al. (2002) derived the local radio
luminosity function for AGN and early-type galaxies of the FIRST-2dF
Galaxy Redshift Survey (2dFGRS), finding that the optical counterparts
have B-band luminosity characterized by a very narrow distribution,
with average $\mb -5 \log h =-19.95$ and scatter $0.41$ mag [$h \equiv
H_0/(100 \km \sm1 \mpc^{-1})$, where $H_0$ is the Hubble constant].

Let us assume that all bright elliptical galaxies have a cooling core,
a central supermassive BH, and that, over the galaxy lifetime, the
radiative losses from their hot plasma are balanced by the mechanical
energy emitted in radio outbursts from the central AGN
($\av{\lmech}=\lx$).  The radio luminosity of the AGN will be strongly
variable, and poorly correlated with the cooling luminosity of the
host cooling core.  However, under the assumption that the
monochromatic 1.4 GHz radio luminosity traces the kinetic luminosity
($\lmech=\k14 \lonefour$), for a sample of galaxies with the same cooling
luminosity $\lx$, we would expect $\av{\lonefour}=\lx/\k14$. If
$\phix(\lx)$ is the distribution of the cooling luminosities of bright
early-type galaxies, the radio luminosity function of early-type
galaxies will be given by
\begin{equation}\label{eqrlf} 
\Phi(\lonefour)= \int\d\lx\, \phix(\lx) \prob(\lonefour;\lx),
\end{equation}
where $\prob(\lonefour;\lx)$ is a log-normal distribution with
$\av{\lonefour}=\lx/\k14$ and variance $\sig14$. The luminosity function
$\phix(\lx)$ has yet to be determined observationally, because of the
lack of complete X-ray samples of early-type galaxies. As well known,
the diffuse X-ray luminosity of early-type galaxies with similar
optical luminosity is characterized by significant scatter (Fabbiano,
Gioia \& Trinchieri 1989; O'Sullivan, Forbes \& Ponman 2001).  Given
that the optical luminosity is roughly the same for all sources, it is
natural to model the distribution of $\lx$ as
\begin{equation}\label{eqxlf}
\Psi(\lx)\d\lx= 
{\frac{N_0}{\sqrt{2\pi}\sigma_{\rm X}}}\,{\exp \left[- {{(
\ell_{\rm X}-{\av{\ell_{\rm X}}})}^2 \over
2\sigx}\right]}\d\ell_{\rm X},
\end{equation}
where $\ell_{\rm X}=\ln(\lx)$, $\sigx$ is the logarithmic variance and
$N_0$ is the number density.  Substituting $\Psi(\lx)$ and
$\prob(\lonefour;\lx)$ in equation~(\ref{eqrlf}) we get
\begin{equation}\label{eqrlfln}
\Phi(\lonefour)\d\lonefour=
{\frac{N_0}{\sqrt{2\pi}\sigma_{\star}}}\,
\exp{\left[-{{(\ell_{1.4}-\ln{L_{1.4}^{\star}})}^2 \over
{2 \sigstar }}\right]}\d\ell_{1.4},
\end{equation}
where $\ell_{1.4}=\ln(\lonefour)$, $\sigstar=\sig14+\sigx$, and
$\ln{L_{1.4}^{\star}}=\av{\ell_{\rm X}}-\ln{\k14}-\sig14/2$.  Thus,
our model predicts that the radio luminosity function of early-type
galaxies is log-normal with variance $\sig14+\sigx$, where $\sig14$ is
the variance due to radio-variability of each source and $\sigx$ is
the variance in the distribution of X-ray emission of bright
ellipticals.

We test this prediction by fitting the local radio luminosity function
of early-type galaxies of the FIRST-2dFGRS (Magliocchetti et al.~2002)
with equation~(\ref{eqrlfln}), leaving $\sigstar$, $L_{1.4}^{\star}$
and $N_0$ as free parameters.  We found best--fitting parameters (with
1~$\sigma$ uncertainties) $\sigstar=7.3\pm{1.0}$,
${L_{1.4}^{\star}}~=~10^{36.5\pm{0.2}}~\ergs~\sm1$, and
$N_0=(1.2\pm{0.3})\times10^{-3} h^3 \mpc^{-3}$. Fig.~7 plots the data
and error bars as reported by Magliocchetti et al. (2002) together
with our best--fitting log-normal function. We note that the best--fit
gives $\sig14+\sigx \simeq 7.3$, consistent with values of $\sig14$
found analysing B04's sample of central radio galaxies in cooling
cores (Section~3.2) plus a contribution from intrinsic scatter in
X-ray luminosity.

We based our analysis of the radio luminosity function on the
assumption that most bright elliptical and S0 galaxies have
AGN-powered radio emission. We can check this assumption a posteriori
by comparing the best--fitting number density $N_0$ with the number
density of luminous early-type galaxies.  We assume that the B-band
optical luminosity function of early-type galaxies $\phi(\lb)$ is a
Schechter function with slope $\alpha$, characteristic luminosity
$\lbstar$, and normalization $\phi^{\star}$.  Then the number density
of objects brighter than $\lb$ is $N_{\rm B}(>\lb)=\int_{\lb}^{\infty}
\d \lb'\,{\phi}(\lb')=\phi^{\star}\Gamma(\alpha+1,\lb/\lbstar)$.  If
we consider the best--fitting B-band luminosity function determined by
Madgwick et al. (2002) for the early-type galaxies of the 2dFGRS
[$\alpha=-0.54\pm{0.02}$, $\phi^{\star}=(9.9\pm0.5)\times10^{-3} h^3
\mpc^{-3}$], and we assume that all early-type galaxies with
$\lb>\lbstar$ are radio galaxies, we find a number density $N_{\rm
B}(>\lbstar)= (2.7\pm0.1)\times10^{-3} h^3 \mpc^{-3}$.  Thus, $N_{\rm
B}(>\lbstar)$ is a factor of $\sim 2$ larger than $N_0$ found from
our best--fitting log-normal distribution.  The discrepancy between
the two number densities is easily explained by our arbitrary choice
of $\lbstar$ as the minimum B-band luminosity of a radio galaxy, and
by uncertainties on the faint-end of radio luminosity function.

We conclude that available radio data are consistent with the proposed
scenario, in which each bright elliptical oscillates, over its
lifetime, along the radio luminosity function, spanning several orders
of magnitude in radio power. Interestingly, results of
high-redshift radio surveys also suggest that most massive ellipticals have
experienced one or more episodes of powerful radio activity during
their lifetime (Rawlings 2003). Finally, we stress that the proposed
picture naturally explains the very large scatter in radio luminosity
for radio galaxies with the same BH mass (Lacy et al. 2001; McLure et
al. 2004, and references therein) as a consequence of the strong
time-variability of their radio emission (see also Nipoti et al.~2005).

\section{Summary and conclusions}

Since the cosmic density of matter is very inhomogeneous on small scales, it
is not possible to derive its global properties from individual measurements
of the density in small regions: the characteristics of large-scale
structure must be determined from a model of the fluctuating density field,
and the values that one derives from it of statistical quantities, such as
the mass variance.  Similarly, the temporal variability of AGN is too large
to ignore. We cannot hope to understand the radio luminosity function, or to
test the AGN/cooling-flow model, without a model of AGN variability.

The normalized mechanical luminosities of a sample of AGN,
$\nlm=\lmech/\lx$, should show scatter only because of temporal variability,
and be free of differences in the scales of systems. Since the variability
of AGN is known to be large and $\nlm$ is an inherently positive variable,
the natural variable to model is $\ell=\ln(\nlm)$, and the default
assumption to make is that $\ell(t)$ is a Gaussian random process. This
assumption implies that individual measures of $\nlm$ should have a
log-normal distribution, so the time-sequence of $\lmech$ is completely
characterized by the power spectrum $P(\omega)$ of $\ell(t)$.

The enormous range of physical scales involved in AGN suggests that the
power spectrum will be a power law $P(\omega)\sim\omega^{-\beta}$. Several
considerations suggest that $\beta\simeq1$, the index characteristic of
flicker noise. First, on the short time-scales that are directly accessible,
monitoring of AGN yields $\beta\simeq1$. Second, a comparison of the
amplitudes measured in these studies with that inferred from the short duty
cycle of quasars requires $\beta\simeq1$. Third, microquasars, whose power
spectra can be probed over a camparatively wide range of frequencies, are
found to have $\beta\simeq1$. Finally, diverse physical phenomena are found
to be characterized by flicker-noise power spectra, so it would not be
surprising if AGN conformed to this pattern.

We estimate the spectral amplitude of the mechanical luminosity of AGN at
the centres of cooling flows by assuming that energy radiated by the thermal
plasma is on the average replaced by the mechanical input of the central
AGN, and that X-ray cavities probe the mechanical luminosity averaged over
time-scales $\sim 10^8\yr$.  The amplitudes required are slightly larger than
those that characterize the radio output of two microquasars, and of the
order of those estimated from the duty cycle of quasars. 

We estimate the coefficients $\kappa$ in an assumed proportionality
between the mechanical luminosity of an AGN and its radio power,
either at $1.4\,$GHz or bolometric. We do this in two ways. In the
first we directly fit a straight line to the correlation between radio
and mechanical luminosities of sources that have radio-loud
cavities. Alternatively, we fit our model of AGN variability to the
radio data of all sources with cavities, whether radio-loud or ghost,
under the assumption that $\av{\lmech}=\lx$.  The two methods yield
compatible ranges of values of $\kappa$.

The available observations fail to constrain the
model's parameters strongly, both because the currently available sample of
sources is small and because our understanding of the physics of these
sources is poor. Despite the significant advance that was made possible by
the discovery of numbers of X-ray cavities, measurements of the ratio of
radio to mechanical luminosity are still extremely uncertain. 
Also, our analysis clearly shows the
importance of determining the duty cycle of the radio sources in order
to quantify their contribution in the energetics of cooling flows: in
particular, it is crucial to estimate the amplitude of their power
spectrum on time-scales $\gsim 10 \myr$.  We conclude that larger
samples of AGN/cooling-flow pairs and better theoretical models of
radio sources are needed to challenge the AGN/cooling-flow model.

In a scenario in which AGN radio activity is strongly variable with
time, the local luminosity function of radio galaxies should be
interpreted as the distribution of the time spent by a galaxy at a
given radio luminosity rather than a distribution of sources with
different intrinsic properties. This picture is supported by the
observational finding that early-type galaxies hosting supermassive
BHs with the same mass span several orders of magnitude in radio
power. We showed that the FIRST-2dFGRS luminosity function of radio
galaxies can be fitted by a log-normal distribution, consistent with
the hypothesis that in all early-type galaxies, the cooling flow is
balanced by a strongly variable radio source. Our model predicts a
turnover at luminosities similar to the lowest currently probed. Any
turnover in the luminosity function of AGN will be to a large extent
masked by radio emission from nuclear starbursts. Nonetheless, the
prediction of a turnover is interesting and will be tested by
forthcoming radio surveys with instruments such as the Square
Kilometer Array.

\section*{Acknowledgments}

We are grateful to Katherine~M.~Blundell and Steve~Rawlings for
helpful discussions. We used data made publicly available by the
GBI-NASA monitoring program.  The Green Bank Interferometer (GBI) was
a facility of the National Science Foundation operated by the National
Radio Astronomy Observatory in support of NASA High Energy
Astrophysics programs.



\bsp

\label{lastpage}

\end{document}